\documentclass[usenatbib]{mn2e}

\usepackage{txfonts}
\usepackage{graphicx}
\usepackage{natbib}

\def\nat{Nature}
\def\apj{ApJ}
\def\mnras{MNRAS}

\def\aap{A\&A}                   
\def\aaps{A\&AS}                 
\def\aapr{A\&A Rev}          
\def\apjs{ApJS}                  
\def\apjl{ApJ}                   

\def\aj{AJ}

\def\Msun{M$_{\odot}$}

\begin{document}

\author[E.~Kuiper et al.]{E.~Kuiper,$^1$\thanks{E--mail: kuiper@strw.leidenuniv.nl} N.~A.~Hatch,$^{2}$ B.~P.~Venemans$^3$, G.~K.~Miley,$^1$ H.~J.~A.~R\"{o}ttgering,$^1$ J.~D.~Kurk,$^4$
\newauthor R.~A.~Overzier,$^5$ L.~Pentericci,$^6$ J.~Bland-Hawthorn,$^{7,8}$ J.~Cepa$^{9,10}$
\\
$^1$ Leiden Observatory, Leiden University, P.O. 9513, Leiden 2300 RA, the Netherlands \\
$^2$ School of Physics and Astronomy, The University of Nottingham, University Park, Nottingham NG7 2RD, UK \\
$^3$ European Southern Observatory, Karl--Schwarzschild Strasse, 85748 Garching bei M\"{unchen}, Germany \\
$^4$ Max--Planck--Institut f\"{u}r Extraterrestrische Physik, Giessenbachstrasse, D-85741 Garching, Germany \\
$^5$ Max--Planck--Institut f\"{u}r Astrophysik, Karl--Schwarzschild Strasse 1, D--85741 Garching, Germany \\
$^6$ INAF, Osservatorio Astronomica di Roma, Via Frascati 33, 00040 Monteporzio, Italy \\
$^7$ Sydney Institute for Astronomy, School of Physics, University of Sydney, NSW 2006, Australia \\
$^8$ Leverhulme Visiting Professor, and Merton College Fellow, University of Oxford, OX1 3RH, UK \\
$^9$ Instituto de Astrof\'{i}sica de Canarias, 38205 La Laguna, Spain \\
$^{10}$ Departamento de Astrof\'{i}sica, Universidad de La Laguna, Spain \\
}

\title[Protoclusters with tunable filters]{Discovery of a high-$z$ protocluster with tunable filters: the case of 6C0140+326 at $z=4.4$}

\maketitle

\begin{abstract}
We present the first results obtained using a tunable narrowband filter in the search for high-$z$ protoclusters. Using the recently commissioned red tunable filter on the Gran Telescopio Canarias we have searched for Ly$\alpha$ emitters in a 75~arcmin$^2$ field centered on the $z=4.413$ radio galaxy 6C0140+326. With three different wavelength tunings we find a total of 27 unique candidate Ly$\alpha$ emitters. The availability of three different wavelength tunings allows us to make estimates of the redshifts for each of the objects. It also allows us to separate a possible protocluster from structure in the immediate foreground. This division shows that the foreground region contains significantly fewer Ly$\alpha$ emitters. Also, the spatial distribution of the objects in the protocluster field deviates from a random distribution at the $2.5\sigma$ level. The observed redshift distribution of the emitters is different from the expected distribution of a blank field at the $\sim3\sigma$ level, with the Ly$\alpha$ emitters concentrated near the radio galaxy at $z>4.38$. The 6C0140+326 field is denser by a factor of $9\pm5$ than a blank field, and the number density of Ly$\alpha$ emitters close to the radio galaxy is similar to that of the $z\sim4.1$ protocluster around TN~J1338-1942. We thus conclude that there is an overdensity of Ly$\alpha$ emitters around the radio galaxy 6C0140+326. This is one of few known overdensities at such a high redshift.
\end{abstract}

\section{Introduction} \label{sec:intro}

The identification of the progenitors of local galaxy clusters at $z>2$ is a difficult task. For the interval $1<z<1.5$, galaxy clusters are most often identified by infrared red sequence searches or observations of the X-ray emitting intracluster gas and the number of galaxy clusters at these redshifts is growing steadily \citep[e.g.][]{stanford1997,rosati1999,rosati2004,mullis2005,stanford2006,muzzin2009,brodwin2010,bielby2010}. Unfortunately, these methods become increasingly less effective when moving beyond $z=1.5$ as the number of red galaxies decreases and X-ray emission becomes too faint to be easily observed. However, if we wish to understand the role of environment on galaxy evolution and the emergence of large scale structure it is essential to locate and study galaxy clusters at all possible epochs. Recent results have presented the spectroscopic confirmation of galaxy clusters with X-ray emission at $z>1.5$ \citep{wilson2008,kurk2009,papovich2010,tanaka2010,henry2010}, with the current distance record being the galaxy cluster CLJ1449+0856 at $z\sim2.07$ presented by \citet{gobat2011} \citep[but see also][]{andreon2011}. However, this sample of high-$z$ clusters remains small.

Another successful method of identifying galaxy cluster progenitors at $z>2$ is to search for overdensities of line emitting galaxies using narrowband imaging. These searches are often aimed at fields containing high-$z$ radio galaxies \citep[hereafter HzRGs,][]{miley2008}, since these are thought to have large stellar masses of the order of 10$^{11}$ to 10$^{12}$~\Msun~\citep{roccavolmerange2004,seymour2007}. According to hierarchical galaxy formation, the most massive galaxies form in the densest environments. The massive nature of HzRGs thus indicates that these objects may trace overdensities in the early Universe. In recent years many studies have focused on finding galaxy overdensities around HzRGs \citep[e.g.][]{pascarelle1996,knopp1997,pentericci2000,kurk2004b,kurk2004a,overzier2006,venemans2007,overzier2008,kuiper2010,galametz2010,hatch2011}. Since these overdensities show no evidence of X-ray emission (at luminosities $>10^{44}$~erg~s$^{-1}$) it is thought that these are forming clusters, not yet dynamically relaxed \citep{carilli2002,overzier2005}. They are therefore often called 'protoclusters'. 

However, even though the number of spectroscopically confirmed $z>2$ HzRGs approaches 200, the fraction of these that have been studied for the presence of galaxy overdensities remains small. This is partly due to the small number of existing narrowband filters and the fact that the central wavelengths of these existing filters are often based on strong lines at $z=0$, such as [O{\sc iii}]$\lambda5007$. This severely limits the redshifts at which protoclusters can be studied and therefore the absolute number of confirmed protoclusters has remained small. 

In this paper we present the results of a pilot study that utilizes tunable narrowband filters in the search for line emitting galaxies around the HzRG 6C0140+326 at $z=4.413$ \citep{rawlings1996,debreuck2001}. Tunable filters (TFs) allow the user to set the central wavelength and width of the narrowband filter. TFs use two plane parallel transparent plates coated with films of high reflectivity and low absorption. By separating the two plates by a small distance of the order of a $\mu$m--mm a cavity is formed which is resonant at a specific wavelength. Constructive interference at the resonant wavelength then causes all the incident light at that wavelength to be transmitted. Changing the separation between the plates then allows the central wavelength or the width of the filter to be adjusted. TFs therefore alleviate the limitations imposed by a small number of available narrowband filters at fixed wavelengths and are thus ideally suited for searching for protoclusters at a range of redshifts. More information and details concerning tunable filters can be found in \citet{hawthorn1995} and \citet{jones2002}.

Similar studies involving the search for line emitting galaxies around $z\sim1$ quasars have been successfully performed by \citet{baker2001} and \citet{barr2004}. An attempt to use this technique at higher redshifts has led to mixed results. In unpublished work by \citet{swinbank2006} a TF study is presented for two radio-loud quasars located at $z\sim2$ and one radio-quite quasar at $z\sim4.5$. The $z\sim4.5$ field shows evidence for an overdensity, whereas the two $z\sim2$ fields lack depth and do not allow for strong conclusions. With the advent of a TF instrument at a 8-10 meter class telescope, it has now become possible to obtain sufficiently deep data to efficiently search for Ly$\alpha$ emitters in the environments of HzRGs at arbitrary redshifts $z>2$.

The paper is organised as follows: in Sect.~\ref{sec:data} we describe the data, its reduction and the object detection. The sample selection and redshift estimation is treated in Sect.~\ref{sec:results} and we discuss the evidence for the presence of an overdensity in Sect.~\ref{sec:disc}. Finally, conclusions and future outlook are presented in Sect.~\ref{sec:conc}. Throughout this paper we use a standard $\Lambda$CDM cosmology with $H_{\rm 0}=71$~km s$^{-1}$~Mpc$^{-1}$, $\Omega_{\rm M}=0.27$ and $\Omega_{\Lambda}=0.73$. All magnitudes given in this paper are in the AB magnitude system.

\section{Data} \label{sec:data}

The radio galaxy 6C0140+326 (hereafter 6C0140) was observed for a total of 18 hours using the Optical System for Imaging and low Resolution Integrated Spectroscopy instrument \citep[OSIRIS,][]{cepa2000,cepa2003} at the Gran Telescopio Canarias (GTC), La Palma. OSIRIS consist of two $2048\times4096$ pixel Marconi CCDs with a 72 pixel gap between the two CCDs. The observations were done on several dates from September 2010 to January 2011. The $2\times2$ binning mode was used resulting in a pixel scale of  $\sim0.25$\arcsec~pixel$^{-1}$ and a total field-of-view of $\sim$8.7\arcmin$\times$8.6\arcmin. The radio galaxy was positioned near the optical centre, approximately 15\arcsec~from the left edge of CCD 2. The individual exposures have been dithered with offsets of approximately 10-12.5\arcsec~in right ascension and 2-4\arcsec~in declination. The larger offsets in right ascension have been chosen such as to cover the gap of $\sim8$~\arcsec~between the two CCDs without losing the radio galaxy in the gap. A full list of details concerning the observations can be found in Table~\ref{table1}.

Broadband images were obtained in the $r$ and $i$ bands. The narrowband images were obtained at three different central wavelengths $\lambda_{\rm c}$. This was done because the central wavelength of the TFs varies across the field of view approximately as
\begin{equation}
\lambda(r)=\lambda_{\rm c}\left(1-0.0007930r^{2}\right)
\end{equation}
as given in the OSIRIS TF user manual. Here $r$ is the distance to the optical centre of the instrument in arcminutes and $\lambda_{\rm c}$ the wavelength at the optical centre. The maximum FWHM of the TF is 20~\AA, therefore it was necessary to perform several passes with different $\lambda_{\rm c}$ in order to cover the entire redshift range of a possible protocluster. At $z\sim4.4$ the Ly$\alpha$ line is shifted to 6580~\AA, so the central wavelengths were chosen to be 6565~\AA, 6575~\AA~and 6585~\AA~(hereafter TF1, TF2 and TF3 for brevity). Note that the redshift of the radio galaxy ($z=4.413$) indicates that it falls between TF2 and TF3. 

\begin{figure*}
\resizebox{\hsize}{!}{\includegraphics{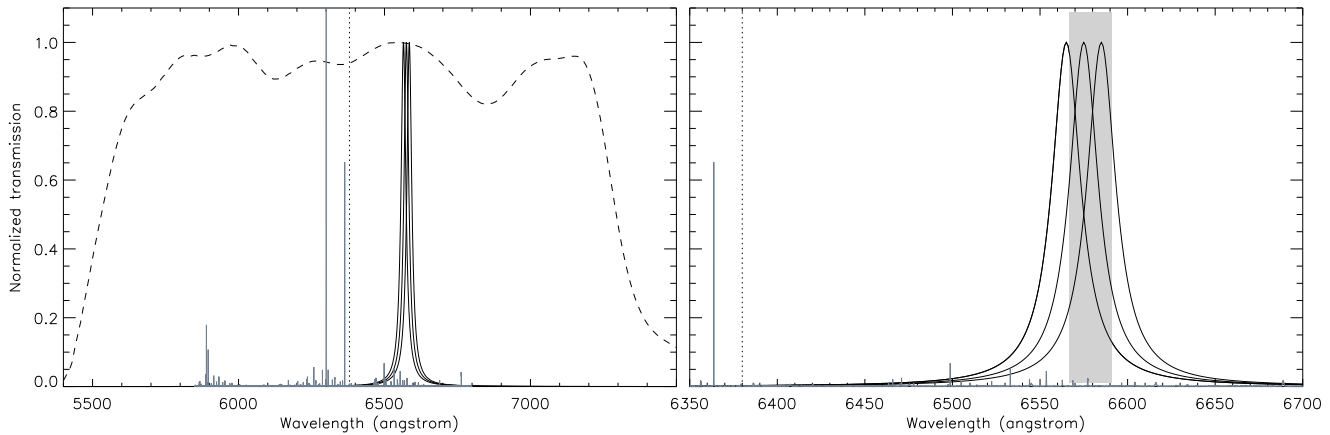}}
\caption{\label{fig:filt}  Left panel: Filter response curves for the three TF tunings (solid curves) and the $r$ band (dashed curve).  Right panel: A zoom in of the three TF response curves in order to show more detail. This panel shows also the approximate area wavelength range where protocluster galaxies are expected to lie. Also shown in grey in both panels is the night--sky emission line spectrum \citep{hanuschik2003}. The minimum wavelength reached at the edge of the field in TF1 is denoted by the vertical dotted line.}
\end{figure*}

As noted above, the maximum formal width of the TF is 20~\AA. However, the shape of the response curve is Lorentzian rather than Gaussian and is approximately given by
\begin{equation} \label{eq:resp}
T=\left\{1+\left[\frac{2(\lambda-\lambda_{\rm c})}{\delta\lambda}\right]^{2}\right\}^{-1}
\end{equation}
with $\lambda_{\rm c}$ the central wavelength and $\delta\lambda$ the formal FWHM. Due to the shape being Lorentzian, the transmission has relatively extended wings which results in an effective band width that is larger than the formal value for the FWHM by a factor of $\pi/2$. Thus the TF tunings have an effective FWHM of $\sim31$~\AA. Taking this into account, our observations probe the redshift range $4.386 < z < 4.428$ near the optical centre. The true redshift range that is covered is larger due to the variation of the central wavelength across the field. The relevant filter response curves are shown in Fig.~\ref{fig:filt} together with a night--sky emission line spectrum. As can be seen, the sky line contamination is relatively mild.

\begin{table*}
\caption{\label{table1} Details of the observations. The 5$\sigma$ limiting magnitudes have been calculated for an aperture diameter of twice the seeing disk. Also note that the values given for $\lambda_{\rm eff}$ for the TF observations are the values at the optical centre.}
\begin{tabular}{c|c|c|c|c|c}
\hline
Band  & Exp. time (sec.) & $\lambda_{\rm eff}$ (\AA)& $\Delta\lambda$ (\AA) & Seeing (arcsec) & 5$\sigma$ limiting magnitude \\
\hline
\hline
$r$  & 2400 & 6417 & 1685 & 0.8 & 25.8 \\
$i$  & 2100 & 7719 & 1483 & 0.8 & 24.8 \\
TF1 & 15680 & 6565 & 31.4 & 0.8 & 25.1 \\
TF2 & 15680 & 6575 & 31.4 & 0.8 &  25.2 \\
TF3 & 13440 & 6585 & 31.4 & 0.8 &  25.1 \\
\hline
\end{tabular}
\end{table*}

\subsection{Data reduction} \label{sec:reduc}

The reduction of all the data is done using {\sc iraf}\footnote[1]{{\sc iraf} is distributed by the National Optical Astronomy Observatory, which is operated by the Association of Universities for Research in Astronomy, Inc., under cooperative agreement with the National Science Foundation.}. The reduction of the broadband images includes the standard steps of bias subtraction and flat-fielding, where the latter is done using sky flats. To remove further large scale gradients a superflat is made from the unregistered science images. This superflat is smoothed and the science frames are subsequently divided by the superflat. The flat science images are then registered with simple offsets in the x and y direction using the {\sc iraf} task {\sc xregister} and co-added together.

The reduction of the TF images follows the same general outline as that for the broadband images with two exceptions. The flat-fielding is done with dome flats and an additional step is included which involves the removal of sky rings. As the central wavelength of the TF filter varies across the field, skylines shift in and out of the filter bandpass causing a pattern of alternating bright and faint concentric rings superimposed on the image. These large scale gradients are of the order 3 to 9 times the rms noise in the individual images. The rings are removed by subtracting a smoothed superflat made using the unregistered science exposures. This superflat is created for each exposure individually as the sky level (and therefore the sky rings) varies between different nights and airmasses. The individual frames are subsequently registered and combined. Finally, all fully reduced science images are registered to the same pixel coordinates using the {\sc iraf} tasks {\sc geomap} and {\sc georegister}. 

Due to the wavelength variation across the field and the dithering there is a variation in wavelength in each of the individual pixels. This variation is larger near the edge of the field of view. The wavelength assigned to each pixel in the final images is the mean of the wavelengths of the pixel in question in the individual images. This also implies that the effective FWHM of the TF increases when moving away from the optical centre. This effect is strongest in the right ascension direction because the dithering steps are larger in this direction, with a maximum increase of $\sim50$~per~cent at the very edges of the images.

Flux calibration for both broad and narrowband imaging is achieved using standard star observations obtained at the end of each observing block of 1 hour. The standard stars used for the TF flux calibration have full SEDs available allowing flux calibration at the exact wavelength of each of the TF observations. A further independent check of the flux calibration of the TF observations is obtained using the broadband data. The $r$ and $i$ magnitudes of all objects in the science frames with $19 < r < 23$ are measured. Then, assuming a power law spectral energy distribution, the magnitudes of these objects at the wavelength of interest are determined. The median zeropoints derived with this method deviate by -0.1, -0.04 and +0.005 magnitude with respect to the standard star zeropoints of TF1, TF2 and TF3. The larger deviation for TF1 and TF2 are likely due to the stronger presence of the H$\alpha$ absorption line at 6563~\AA.

The final reduced and coadded TF images show some artifacts of the reduction. The sky ring subtraction is not optimal due to the applied smoothing and therefore a residual ring pattern remains in the final images. Also the unique properties of the TF lead to pupil ghosts near bright stars. Point source ghosts, however, are not present in the final TF images. The dithering results in an offset in point source ghosts opposite to the actual offset. When combining the individual images the point source ghosts will therefore be removed. 

\subsection{Source detection and photometry}

Object detection and photometry are done using {\sc SExtractor} \citep{bertin1996} in double image mode. We create three different catalogues based on each of the three TF tunings. A detection is defined as a minimum of 9 adjacent pixels that each exceed the 2$\sigma$ rms noise. Colours are measured using the 2$\sigma$ isophotal apertures as determined from the respective detection image whereas total magnitudes are measured using {\sc SExtractor}'s {\sc mag\_auto} apertures. Image depth and uncertainties on the photometry are determined using the method of \citet{labbe2003}.

Completeness of the image is measured by adding point sources of a range of magnitudes to the respective images after which source extraction is repeated and the number of recovered objects is assessed. To avoid overcrowding the image we limit the number of added objects to 150 per magnitude. This process is repeated 10 times in order to obtain better statistics. We find that the data is 50 per cent complete for point sources down to $r$=26.0, TF1=25.1, TF2=25.1 and TF3=25.0 magnitudes.

\section{Results} \label{sec:results}

\subsection{Selection of LAEs} \label{sec:selec}

\begin{figure*}
\resizebox{\hsize}{!}{\includegraphics{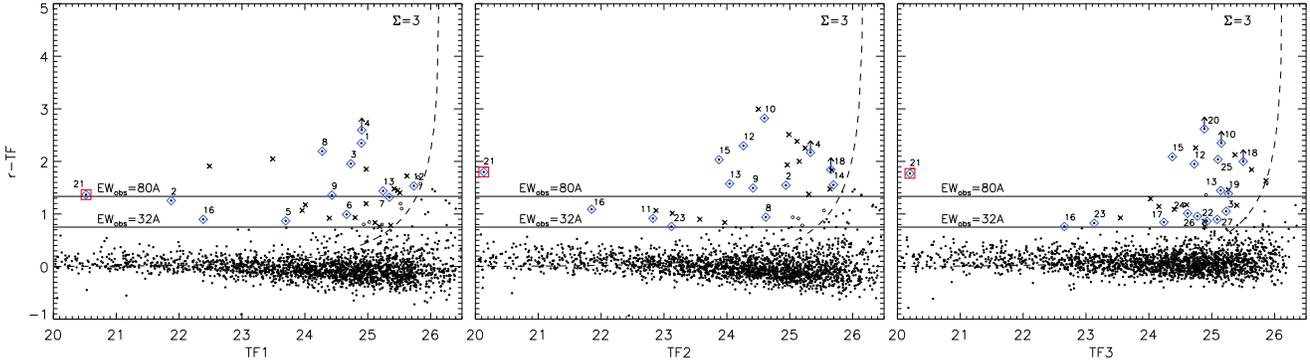}}
\caption{\label{fig:cmds} Colour-magnitude diagrams for each of the three TF tunings. The horizontal solid lines indicate $EW_{\rm obs}$ values of 32 and 80~\AA~respectively, whereas the dashed curve indicates the $\Sigma=3$ line. Objects identified as LAEs are denoted by diamonds and labeled with the respective ID numbers as listed in Table~\ref{table2} and Table~\ref{table3}. The radio galaxy is denoted by the square symbol. The $\Sigma=3$ curve is calculated using a median aperture size. Individual objects may therefore be located inside the selection area as depicted here, but are still not identified as being a LAE. These objects are denoted by open circles. Spurious detections are indicated by crosses.}
\end{figure*}

For each of the three TF tunings a separate sample of Ly$\alpha$ emitters (hereafter LAEs) is selected. The criterium for identifying LAEs is based on the colour-magnitude diagrams of the 6C0140 field as shown in Fig.~\ref{fig:cmds}. The galaxies with line emission at the relevant redshift will show an excess of flux in the TF band relative to the $r$ band flux. Most objects do not have emission lines in the TF and thus have $r-{\rm TF}\sim0$. The rms scatter around $r-{\rm TF}\sim0$ at ${\rm TF}\sim24-25$~mag is $\sim0.15$. For an object to be identified as LAE we require at least a $5\sigma$ deviation from $r-{\rm TF}\sim0$ and thus that $r-{\rm TF}>0.75$. 

This excess flux relates to an observed line equivalent width using the relationship
\begin{equation} \label{eq:ew}
EW_{\rm obs}=\frac{\Delta\lambda_{r}\Delta\lambda_{\rm TF}\left[1-10^{-0.4(r-{\rm TF})}\right]}{\left[\Delta\lambda_{r} 10^{-0.4(r-{\rm TF})}-\Delta\lambda_{\rm TF}\right]}
\end{equation}
from \citet{bunker1995}. Here $\Delta\lambda$ is the FWHM of the filter in question and $r$ and TF are the measured magnitudes in the broad- and narrowband respectively. The restframe equivalent width $EW_{\rm 0}$ is obtained by dividing by $(1+z)$. Note that this relation does not account for IGM absorption. For LAEs at $z\sim4.4$ this implies that $EW_{\rm obs}$ and $EW_{\rm 0}$ are overestimated by a factor of $\sim1.5-2$. We will return to this correction factor in Sect.~\ref{sec:redz}.

Using Eq.~\ref{eq:ew}, the colour cut corresponds to approximately $EW_{\rm obs}>32$~\AA~at the optical centre of the field. However, as described in Sect.~\ref{sec:reduc}, the dithering results in an effective broadening of the TF when moving away from the optical centre. At the West and East edges of the image the increase in effective FWHM is $\sim50$ per cent, thus the equivalent width cut varies approximately between $EW_{\rm obs}>32$~\AA~and $EW_{\rm obs}>48$~\AA~across the field.

The equivalent width cut used in this work is less stringent than the more commonly used $EW_{\rm obs}>80$~\AA. This is because the TF is narrower than conventional narrowband filters, thus allowing for lower equivalent width objects to be included. We further require that the error parameter $\Sigma>3$, i.e. the excess flux is at least three times larger than the combined noise of the measured broad- and narrowband fluxes. Finally, due to the artifacts present in the images we visually inspect all of the objects that satisfy the above criteria and discard any spurious detections.

\begin{figure}
\resizebox{\hsize}{!}{\includegraphics{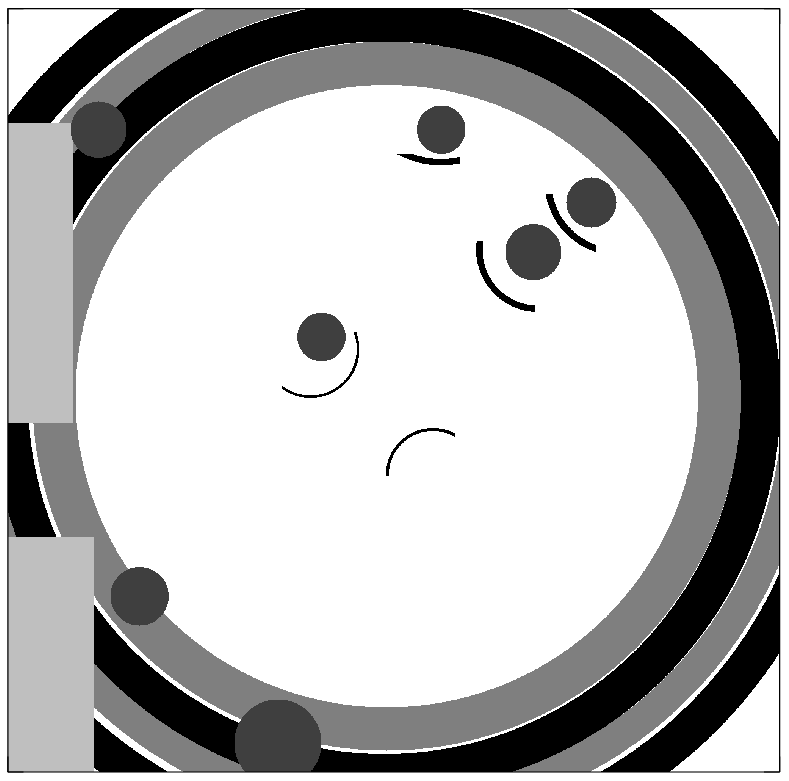}}
\caption{\label{fig:mask} A mask of the 6C0140 field showing the location of regions with large numbers of spurious detections and regions where source detection is impeded. The large grey and black rings are the approximate locations of the sky rings in TF1 and TF3, respectively. For TF2 the sky rings fall in between the two sets shown here. The small black arcs denote the locations of the pupil ghosts, whereas the filled circles denote the location of bright stars. The two light grey regions at the left side of the image are affected by vignetting and therefore not considered.}
\end{figure}

As can be seen from Fig.~\ref{fig:cmds}, there is a significant number of objects located in the selection area that are not identified as being LAEs.  Visual inspection shows that these objects are likely spurious detections because they are often found in the sky ring residuals, have unphysical shapes, coincide with the pupil ghosts or are located near bright stars and saturation spikes. These objects are therefore not included in the candidate LAE sample.  Figure~\ref{fig:mask} shows the regions of the image where most of these spurious are located, i.e. the sky rings and the pupil ghosts. Also shown are the locations of bright stars and portions of the image that are affected by vignetting. The bright stars cover $\sim4$~per~cent of the field and therefore do not influence any of the conclusions presented in this work.

Other objects close in the selection area may actually have $\Sigma<3$. Individual values of $\Sigma$ depend on the aperture size for each object. The $\Sigma=3$ curves shown in Fig.~\ref{fig:cmds}, however, are calculated for a fixed aperture size that is taken to be the median aperture size of the LAEs. Thus individual objects inside the selection area may have $\Sigma<3$ or vice versa.

\begin{table*}
\caption{\label{table2} Coordinates and magnitudes of the detected LAEs. The table is divided in three sections. The objects in the first section are brightest in TF1 and so forth. $^a$The radio galaxy. The $r$ band was used as the detection image for this object.}
\begin{tabular}{lc|c|c|c|c|c|c|c|}
\hline
ID &  RA & Dec. & Detected in & $r$ & TF1 & TF2 & TF3 \\
\hline
\hline
\#1 & 01:43:58.2 & +32:49:49.8 & TF1 & $27.2\pm0.8$ & $24.9\pm0.2$& $25.34\pm0.3$ & $26.3\pm0.6$  \\
\#2 & 01:43:44.6 & +32:49:42.6 & TF1 & $23.1\pm0.1$ & $21.87\pm0.04$ & $22.55\pm0.06$ & $22.9\pm0.1$ \\
\#3 & 01:43:44.8 & +32:57:06.0 & TF1,TF2,TF3 & $26.7\pm0.4$ & $24.7\pm0.1$ & $25.0\pm0.2$ & $25.4\pm0.2$ \\
\#4 & 01:43:38.9 & +32:53:37.2 & TF1,TF2 & $>27.5$ & $24.9\pm0.2$ & $25.2\pm0.2$ & $25.7\pm0.3$  \\
\#5 & 01:43:38.3 & +32:49:58.3 & TF1 & $24.6\pm0.1$ & $23.7\pm0.1$ & $23.9\pm0.1$ & $24.2\pm0.2$ \\
\#6 & 01:43:35.7 & +32:54:36.0 & TF1 & $25.7\pm0.2$ & $24.7\pm0.1$ & $25.1\pm0.2$ & $25.4\pm0.3$ \\
\#7 & 01:43:35.4 & +32:52:11.5 & TF1 & $26.7\pm0.3$ & $25.3\pm0.2$ & $25.8\pm0.2$ & $25.8\pm0.2$ \\
\#8 & 01:43:24.8 & +32:54:09.5 & TF1,TF2 & $26.5\pm0.4$ & $24.3\pm0.1$ & $25.4\pm0.3$ & $25.1\pm0.2$ \\
\hline
\#9 & 01:43:45.9 & +32:53:58.0 & TF1,TF2 & $25.9\pm0.2$ & $24.4\pm0.1$ & $24.4\pm0.1$ & $24.8\pm0.2$ \\
\#10 & 01:43:43.3 & +32:52:08.1 & TF2,TF3 & $27.4\pm1.0$ & $25.5\pm0.3$ & $24.6\pm0.1$ & $24.8\pm0.2$ \\
\#11 & 01:43:42.1 & +32:55:27.6 & TF2 & $23.7\pm0.1$ & $23.1\pm0.1$ & $22.8\pm0.1$ & $23.1\pm0.1$ \\
\#12 & 01:43:41.5 & +32:54:17.2 & TF1,TF2,TF3 & $26.6\pm0.5$ & $25.3\pm0.3$ & $24.3\pm0.1$ & $24.5\pm0.2$ \\
\#13 & 01:43:41.4 & +32:53:49.3 & TF1,TF2,TF3 & $25.6\pm0.3$ & $24.6\pm0.2$ & $24.0\pm0.1$ & $24.4\pm0.1$ \\
\#14 & 01:43:41.0 & +32:53:48.5 & TF2 & $27.2\pm0.5$ & $26.0\pm0.3$ & $25.7\pm0.2$ & $26.3\pm0.4$ \\
\#15 & 01:43:40.2 & +32:55:02.5 & TF2,TF3 & $25.9\pm0.4$ & $24.7\pm0.2$ & $23.9\pm0.1$ & $24.1\pm0.2$  \\
\#16 & 01:43:38.0 & +32:49:51.9 & TF1,TF2,TF3 & $22.9\pm0.1$ & $22.19\pm0.07 $ & $21.85\pm0.05$  & $22.35\pm0.08$  \\
\hline
\#17 & 01:43:59.8 & +32:52:16.2 & TF3 & $25.1\pm0.1$ & $25.0\pm0.2$ & $24.8\pm0.1$ & $24.2\pm0.1$ \\
\#18 & 01:43:56.4 & +32:54:41.1 & TF2,TF3 & $>27.5$ & $26.2\pm0.3$ & $25.7\pm0.2$ & $25.5\pm0.2$ \\
\#19 & 01:43:44.8 & +32:56:01.9 & TF3 & $26.7\pm0.2$ & $26.7\pm0.4$ & $27.0\pm0.5$ & $25.3\pm0.1$ \\
\#20 & 01:43:43.9 & +32:52:28.6 & TF3 & $>27.5$ & $27.5\pm1.6$ & $25.7\pm0.3$ & $24.9\pm0.1$ \\
\#21$^{a}$ & 01:43:43.8 & +32:53:49.9 & TF1,TF2,TF3 & $23.65\pm0.05$ & $21.54\pm0.01$ & $20.91\pm0.01$ & $20.75\pm0.01$\\
\#22 & 01:43:38.0 & +32:52:00.9 & TF3 & $25.8\pm0.1$ & $25.8\pm0.2$ & $25.6\pm0.2$ & $24.9\pm0.1$ \\
\#23 & 01:43:36.1 & +32:55:00.9 & TF2,TF3 & $24.0\pm0.1$ & $23.8\pm0.1$ & $23.17\pm0.05$ & $23.13\pm0.05$ \\  
\#24 & 01:43:33.3 & +32:54.09.0 & TF3 & $25.6\pm0.2$ & $26.1\pm0.5$ & $25.2\pm0.2$ & $24.6\pm0.1$  \\
\#25 & 01:43:27.3 & +32:51:32.4 & TF3 & $27.1\pm0.5$ & $27.5\pm1.4$ & $25.7\pm0.3$ & $25.1\pm0.2$ \\
\#26 & 01:43:26.6 & +32:52:17.3 & TF3 & $25.7\pm0.2$ & $27.1\pm0.9$ & $25.5\pm0.2$ & $24.8\pm0.1$ \\
\#27 & 01:43:25.9 & +32:52:01.8 & TF3 & $26.0\pm0.2$ & $26.4\pm0.6$ & $25.9\pm0.3$ & $25.1\pm0.2$ \\
\hline
\end{tabular}
\end{table*}

For the TF1 filter, which probes exclusively $z<4.407$, we find a total of 13 candidate LAEs, including the radio galaxy which has $EW_{\rm obs}=88.3$~\AA. This is smaller than expected as visual inspection of the TF1 image indicates that the radio galaxy does have a strong narrowband excess. However, the Ly$\alpha$ emission is extended and coincides with a $z\sim0.9$ foreground galaxy \citep{rawlings1996}. This results in strong contamination of the $r$ band flux and therefore the expected narrowband excess is significantly diminished. To alleviate this problem we use the $r$ band to define the colour apertures for the radio galaxy. This results in $EW_{\rm obs}=217$~\AA~indicating that the presence of the foreground galaxy is indeed important.

The TF2 band at 6575~\AA~reveals a total of 14 candidate LAEs, including the radio galaxy with $EW_{\rm obs}=470$~\AA. Eight of the LAEs have been identified in TF1 as well.

Finally, a total of 17 objects are identified as being candidate LAEs in the TF3 band at 6585~\AA. Again, the radio galaxy has been included and its equivalent width is highest in TF3 with $EW_{\rm obs}=578$~\AA. Out of the remaining galaxies eight have not been detected in TF1 or TF2, thus yielding a total number of 27 unique LAEs (including the radio galaxy) in the field of 6C0140. The properties of the emitters are listed in Table~\ref{table2} and Table~\ref{table3}.

\subsection{Redshift distribution} \label{sec:redz}

Having the multiple tunings of the TF means the redshift of the LAEs can be constrained to a greater accuracy than with a single narrowband image. This allows us to investigate the approximate redshift distribution of the candidate LAEs in the field.

In order to obtain the best possible estimated redshifts we model what the effect is of a variety of weighting schemes. We model a simple flat spectrum ($\beta=-2$) with an emission line with restframe FWHM of 250~km~s$^{-1}$ or 500~km~s$^{-1}$ at certain wavelengths in the range $6555<\lambda<6595$~\AA. These values for the FWHM are consistent with the values found for 80 to 90 per cent of LAEs around HzRGs \citep{venemans2005,venemans2007}. Absorption by the IGM is taken into account using the \citet{madau1995} recipe. The modeled spectra are then convolved with the TF response curve as given in Eq.~\ref{eq:resp} with $\lambda_{\rm c}=6565, 6575$ and 6585~\AA. Using the fluxes in each of the TF tunings we then derive estimates for the wavelength of the emission line.

\begin{figure}
\resizebox{\hsize}{!}{\includegraphics{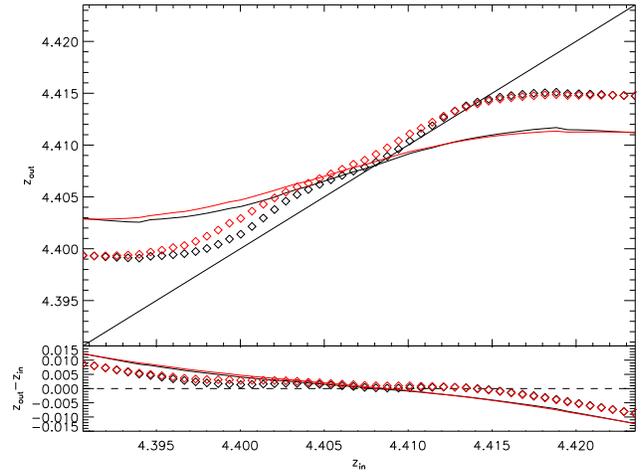}}
\caption{\label{fig:zmod} Upper panel: Output versus input redshift when using the two different weighting schemes tested on modeled Ly$\alpha$ spectral lines. The solid curves indicate the results obtained when using $w_{i}=1$, whereas the diamonds denote the results obtained with the weights described as in Eq.~\ref{eq:weights}. Black and red indicate restframe FWHM of 250~km~s$^{-1}$ and 500~km~s$^{-1}$, respectively. Lower panel: Difference between the output and input redshifts.}
\end{figure}

The results of this process are shown in Fig.~\ref{fig:zmod}. We show two different weighting schemes which can be given in general form as
\begin{equation}
\lambda_{\rm eff}=\frac{\Sigma_{i=1,3}{w_{i}}F_{i}\lambda_{i}}{\Sigma_{i=1,3}w_{i}F_{i}}.
\end{equation}
Here $i$ is TF1, TF2 or TF3 respectively, $F_i$ is the flux in the respective bands, $\lambda_{i}$ is the central wavelength of the tuning in question at the relevant location in the field and $w_{i}$ is a weighting factor. The black data points represent the case of $w_{i}=1$ for all bands and the red data points denote the results obtained for 
\begin{equation} \label{eq:weights}
w_{i}=2^{F_{i}/{\rm min}(F_{\rm TF1},F_{\rm TF2},F_{\rm TF3})}.
\end{equation}
As noted above, the results shown in Fig.~\ref{fig:zmod} are obtained when using the central wavelength values of the optical centre, i.e. $\lambda_{\rm TF1}=6565$~\AA, $\lambda_{\rm TF2}=6575$~\AA~and $\lambda_{\rm TF3}=6585$~\AA. However, the wavelength shift acts on each tuning identically and thus the qualitative behaviour shown in Fig.~\ref{fig:zmod} is valid for the entire field, irrespective of wavelength shift.

We see that the more involved weighting scheme yields better results for almost all redshifts. However, for both weighting schemes the largest discrepancy between the input and output redshifts is at either end of the investigated redshift range. This is to be expected, because no data is available to bracket the existing tunings, hence skewing the output redshift towards a central value. The more elaborate weighting scheme alleviates this slightly, but does not yield full agreement. Figure~\ref{fig:zmod} also shows that the results of the two weightings schemes are fairly robust with respect to the choice of restframe FWHM.

The redshifts obtained when using the weights as described in Eq.~\ref{eq:weights} are listed in Table~\ref{table3}. Uncertainties have been calculated by varying the measured fluxes according to their respective uncertainties and recalculating $\lambda_{\rm eff}$. An additional systematic uncertainty was added in quadrature to take into account the limitations of the weighting scheme. This systematic uncertainty is based on the bands in which the object is detected and the offset from the input redshift implied by this as measured from Fig.~\ref{fig:zmod}. Finally, to account for the variation in wavelength in each pixel we add an additional location-dependent uncertainty in quadrature.

\begin{table}
\begin{center}
\caption{\label{table3} Properties of the detected LAEs. $^a$$EW_{\rm obs}$ calculated using Eq.~\ref{eq:ew}, whereas $EW_{\rm 0}$ is calculated using the method of \citet{venemans2005}. $^b$$EW_{\rm obs}$, $EW_{\rm 0}$, $\Sigma$ and $F_{\rm Ly\alpha}$ for the radio galaxy are calculated on the basis of TF3. The horizontal dividers are as in Table~\ref{table2}.}
\begin{tabular}{lc|c|c|c|c}
\hline
ID & $EW_{\rm obs}/EW_{\rm 0}$ (\AA)$^{a}$ & $\Sigma$ & $F_{\rm Ly\alpha}$ (erg~s$^{-1}$~cm$^{-2}$) & $z_{\rm eff}$\\
\hline
\hline
\#1  & 411/36.2 & 5.2 & $1.4\times10^{-17}$ & 4.310$^{+0.005}_{-0.009}$ \\
\#2  & 84.5/9.1 & 17.9 & $9.0\times10^{-17}$ & 4.337$^{+0.002}_{-0.008}$ \\
\#3 & 209/24.7 & 6.8 & $1.7\times10^{-17}$ & 4.352$^{+0.003}_{-0.007}$ \\
\#4 & $>463/>49.8$ & 6.2 & $1.7\times10^{-17}$ & 4.394$^{+0.005}_{-0.006}$ \\
\#5 & 47.6/8.0 & 5.0 & $2.3\times10^{-17}$ & 4.340$^{+0.004}_{-0.009}$ \\
\#6 & 60.3/6.5 & 4.8 & $1.5\times10^{-17}$ & 4.380$^{+0.005}_{-0.008}$ \\
\#7 & 96.9/13.0 & 4.7 & $1.8\times10^{-17}$ & 4.377$^{+0.005}_{-0.009}$ \\
\#8 & 266/29.6 & 8.9 & $2.2\times10^{-17}$ & 4.327$^{+0.002}_{-0.008}$ \\
\hline
\#9 & 106/11.7 & 6.7 & $1.9\times10^{-17}$ & 4.404$^{+0.002}_{-0.004}$ \\
\#10 & 552/61.5 & 6.9 & $1.5\times10^{-17}$ & 4.399$^{+0.004}_{-0.002}$ \\
\#11 & 47.5/5.0 & 10.3 & $3.7\times10^{-17}$ & 4.391$^{+0.001}_{-0.002}$ \\
\#12 & 299/31.0 & 7.4 & $2.5\times10^{-17}$ & 4.404$^{+0.004}_{-0.002}$ \\
\#13 & 122/12.9 & 7.2 & $3.7\times10^{-17}$ & 4.404$^{+0.002}_{-0.002}$ \\
\#14 & 120/12.4 & 3.5 & $7.0\times10^{-18}$ & 4.403$^{+0.003}_{-0.006}$ \\
\#15 & 225/23.5 & 7.8 & $4.3\times10^{-17}$ & 4.395$^{+0.003}_{-0.002}$ \\
\#16 & 68.8/7.1 & 12.3 & $1.1\times10^{-16}$ & 4.340$^{+0.003}_{-0.003}$ \\
\hline
\#17 & 52.0/4.5 & 5.8 & $1.1\times10^{-17}$ & 4.368$^{+0.009}_{-0.005}$ \\
\#18 & $>247/>25.9$ & 5.0 & $1.1\times10^{-17}$ & 4.385$^{+0.006}_{-0.006}$ \\
\#19 & 96.8/10.0 & 6.0 & $5.5\times10^{-18}$ & 4.387$^{+0.009}_{-0.004}$ \\
\#20 & $>434/>45.0$ & 6.4 & $1.9\times10^{-17}$ & 4.409$^{+0.008}_{-0.001}$ \\
\#21$^{b}$ & 578/75.0 & 95.5 & $3.6\times10^{-15}$ & 4.412$^{+0.001}_{-0.001}$ \\
\#22 & 47.3/4.5 & 5.0 & $6.9\times10^{-18}$ & 4.394$^{+0.008}_{-0.003}$ \\
\#23 & 45.6/4.6 & 8.7 & $2.8\times10^{-17}$ & 4.388$^{+0.003}_{-0.003}$ \\
\#24 & 65.5/5.9 & 4.8 & $1.6\times10^{-17}$  & 4.388$^{+0.008}_{-0.004}$ \\
\#25 & 300/24.3 & 5.2 & $1.8\times10^{-17}$ & 4.339$^{+0.009}_{-0.005}$ \\
\#26 & 68.2/5.8 & 5.3 & $9.6\times10^{-18}$ & 4.344$^{+0.009}_{-0.006}$\\
\#27 & 57.8/5.3 & 3.2 & $1.0\times10^{-17}$ & 4.339$^{+0.009}_{-0.005}$ \\
\hline
\end{tabular}
\end{center}
\end{table}

None of the LAEs are located at larger redshifts than that of the radio galaxy. This is a selection effect as the highest value of $\lambda_{\rm c}$ is 6585~\AA~whereas the Ly$\alpha$ line for $z_{\rm RG}=4.413$ falls at $\sim6580$~\AA. Combined with the wavelength shift towards shorter wavelengths across the field this means that the observations are biased towards redshifts lower than the redshift of the radio galaxy. Also, note that the estimated redshift of the radio galaxy is only marginally inconsistent with the spectroscopic redshift of $z=4.413$. The slight underestimate with respect to the spectroscopic redshift is possibly due to the fact that we modeled the Ly$\alpha$ line with a maximum FWHM of 500~km~s$^{-1}$, whereas the line is observed to have a FWHM of $\sim1500$~km~s$^{-1}$ \citep{rawlings1996,debreuck2001}. A larger line width will introduce stronger systematic uncertainties that have not been taken into account in the case of the estimated redshift of the radio galaxy. This may therefore account for the discrepancy.

The redshift estimates can be used to correct the equivalent width values for IGM absorption. As discussed in Sect.~\ref{sec:selec}, Eq.~\ref{eq:ew} does not take this into account and underestimates the $r$ continuum flux density. Therefore $EW_{\rm obs}$ and $EW_{\rm 0}$ are overestimated. The corrections are calculated following the method of \citet{venemans2005}. The resulting corrected restframe equivalent widths and corresponding Ly$\alpha$ fluxes are listed in Table~\ref{table3}. The difference between $EW_{\rm obs}$ and $EW_{\rm 0}$ is typically a factor $\sim10$. This is consistent with a factor $(1+z)$ in combination with a factor $\sim1.7$. Here the latter factor originates from the fact that a larger portion of the $r$ band flux is absorbed by the IGM compared to the TF flux. The exact factor varies between $\sim1-2$ and depends on the location of the line within the filters and therefore both the redshift of the object and its position in the field.

\subsection{Contamination} \label{sec:interlopers}

One of the larger caveats of using narrowband imaging to select high-$z$ emission line galaxies is that the final sample may be contaminated by low-$z$ interlopers that have a strong emission line falling in the narrowband. For our study the most likely interlopers are [O{\sc ii}] emitters at $z\sim0.76$ or [O{\sc iii}] emitters at $z\sim0.3$. Spectroscopic follow-up is needed to accurately determine the success rate of the sample presented in this work. However, based on previous spectroscopic studies of $z\sim4$ narrowband surveys the expected number of interlopers in our sample can be estimated and it can be determined whether the results presented here are  robust when this is taken into account.

We base our estimate of the success rate on the studies of HzRG TN~J1338-1942 (hereafter 1338) by \citet{venemans2002,venemans2007}, the field study at $z\sim4.5$ by \citet{dawson2007} and the study of the LAEs around $z\sim5.2$ HzRG TN~J0924-2201 by \citet{venemans2004}. The success rates in each of these works are fairly similar to each other, ranging from $\sim75$ per cent to $\sim95$ depending on whether non-detections are counted as non-confirmations. To investigate the `worst case scenario' we use for our sample the minimum success rate of 75 per cent. 

\section{Does 6C0140+326 reside in a protocluster?}  \label{sec:disc}

Based on our sample of LAEs we determine whether there is an overdensity around 6C0140. Due to the wavelength shift across the field, part of the observed field can act as a control field. 

\begin{figure}
\resizebox{\hsize}{!}{\includegraphics{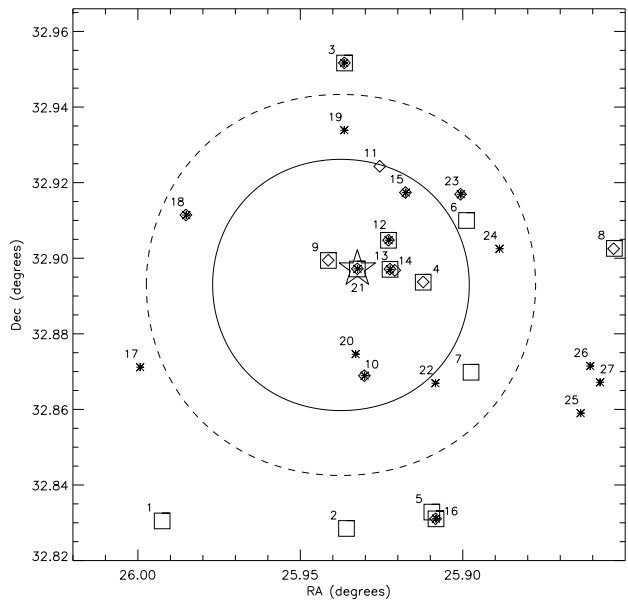}}
\caption{\label{fig:spatial} The spatial distribution of the LAEs. Objects detected in TF1, TF2 and TF3 are marked by squares, diamonds and asterisks respectively. Each object is also marked by its ID number. The location of the radio galaxy is marked by a star. Also shown are two circles denoting the border of a protocluster field (within the full circle) and a foreground field (between dashed and full circle). }
\end{figure}

In Fig.~\ref{fig:spatial} the spatial distribution of the LAEs is shown. Also shown are two concentric circles indicating the boundaries of two fields: the inner circle of $\sim12.5$~arcmin$^{2}$ and the annulus of $\sim16.3$~arcmin$^{2}$, respectively. The central field covers the redshifts closest to the radio galaxy and can therefore be considered a possible protocluster field. The annulus outside this field delimited by the dashed circle probes lower redshifts and is considered to be field environment. To make the distinction between protocluster and foreground stronger we will only consider TF2 and TF3 detected objects (diamonds and asterisks) in the central field, whereas in the annulus only TF1 detected objects are considered (squares). This selection effectively means we are limited to $z>4.38$ in the central field and $z<4.38$ in the annulus. Furthermore, the width of the TF does not change significantly across the central field and the annulus. Since we only consider TF1 in the annulus, the physical depth of the annulus is thus $\sim1.3$ times smaller. We thus require the area of the annulus to be larger by the same factor to have the same volume in each of the fields. We find 9 objects (excluding the radio galaxy) in the possible protocluster field versus 2 in the foreground field. This thus indicates that there is a concentration of LAEs near to the radio galaxy.

It is striking that almost all of the objects within the protocluster field are located west of the radio galaxy in a North-South filamentary structure. We test whether the spatial distribution is consistent with a random distribution by applying a two dimensional Kolmogorov-Smirnov test. There is a probability of 0.01 that the distribution as shown in Fig.~\ref{fig:spatial} is drawn from a random distribution. The distribution is thus different from random at the $\sim2.5\sigma$ level. This further indicates that the LAEs are clustered.

Since our control field is not very large, it is susceptible to cosmic variance. To better quantify the overdensity of LAEs around the radio galaxy, we also compare it to the blank field LAEs observed by \citet{dawson2007}. \citet{dawson2007} presented a differential Ly$\alpha$ luminosity function for field LAEs with $EW_{\rm obs}>80$~\AA~at $z\sim4.5$. Fitting a Schechter function to the luminosity function using a fixed value of $\alpha=-1.6$ they find $\Phi^{\star}=(1.7\pm0.2)\times10^{-4}$~Mpc$^{-3}$ and $L^{\star}=(10.9\pm3.3)\times10^{42}$~erg~s$^{-1}$. 

For the 6C0140 field, if we take into account the overlap between the different tunings, the total `unique' volume probed by the protocluster field is 1570~Mpc$^{3}$. Using the same selection procedure as in \citet{dawson2007}, we find eight emitters (excluding the radio galaxy) in the central field. We calculate the expected number of LAEs in the same field (assuming it is a blank field) using the field luminosity function. The expected number of LAEs is found to be $0.9^{+0.4}_{-0.3}$, with the uncertainty derived from the uncertainties on $\Phi^{\star}$ and $L^{\star}$. Here we use the Ly$\alpha$ flux of the faintest emitter with $EW_{\rm obs}>80$~\AA~(\#14) in the protocluster field as lower limit \citep[$7.0\times10^{-18}$~erg~s$^{-1}$~cm$^{-2}$, this equals $L=1.4\times10^{42}$~erg~s$^{-1}$ for the cosmology used by][]{dawson2007}. We therefore find that the 6C0140 field is denser than a blank field by a factor $9\pm5$, where the uncertainty is based on Poisson statistics and the uncertainty on the expected number. Defining galaxy overdensity as $\delta_{\rm g}=n_{\rm cluster}/n_{\rm field}-1$, we thus find a galaxy overdensity of $8\pm5$. 

Using the overdensity in 6C0140 protocluster field we can make a rough estimate of the mass contained in this central field. As in \citet{venemans2005} we use the relation
\begin{equation}
M=\bar{\rho} V \left(1+\frac{\delta_{\rm g}}{b} \right)
\end{equation}
with $\bar{\rho}$ the mean density of the Universe, $V$ the comoving volume considered and $b$ the bias parameter which relates the galaxy overdensity to the matter overdensity. Following \citet{steidel1998} and \citet{shimasaku2003} we use $b=3-6$. Using $\bar{\rho}=3.5\times10^{10}$~\Msun~Mpc$^{-3}$ and $V=1570$~Mpc$^{3}$ a mass of $0.8-2.9\times10^{14}$~\Msun~is found. This is a strict lower limit to the mass of the entire overdensity because the true extent of the protocluster is likely larger than what is indicated by the central protocluster field.

We also determine whether the number density and redshift distribution of LAEs in the 6C0140 field are consistent with that of a $z\sim4$ protocluster. However, the number of known protoclusters above $z=4$ is limited. One of the few spectroscopically confirmed cases is the protocluster around 1338 at $z\sim4.1$. \citet{venemans2002} and \citet{venemans2007} have shown that the field around the radio galaxy is denser in LAEs than a blank field by a factor of 4.8$^{+1.1}_{-0.8}$. Also, \citet{overzier2006} provided evidence for a relatively large number of Lyman Break Galaxies (LBGs) in this field.

\citet{venemans2007} found a total of 54 LAEs with $EW_{\rm 0}>15$~\AA~in a field of 79.7~arcmin$^{2}$ around 1338. The narrowband filter used was a custom filter with $\lambda_{\rm c}=6199$~\AA~and a FWHM of 59~\AA~(hereafter NB620 for brevity), i.e. approximately twice as wide as the TF tunings used in this study. This difference in width implies that the same emission line, at the respective proper redshifts, will yield a brighter magnitude in the TF. Applying the 50 per cent magnitude limits to the 1338 catalogue will therefore not yield a proper comparison. The recovered Ly$\alpha$ flux is, however, relatively independent of filter width. We therefore use a cut of $F_{\rm Ly\alpha}>7.0\times10^{-18}$~erg~s$^{-1}$~cm$^{-2}$. At $z\sim4.1$ the width of NB620 implies a comoving volume of 12292~Mpc$^3$. Applying the $F_{\rm Ly\alpha}$ cut the density of LAEs in the 1338 field is found to be $3.6\times10^{-3}$~Mpc$^{-3}$. 

Applying the \citet{venemans2007} selection criteria and excluding the radio galaxy, we find a total of 5 unique LAEs with $EW_{\rm 0}>15$~\AA~in the central protocluster field. With a volume of 1570~Mpc$^{3}$ the density of candidate LAEs is thus $(3.2\pm1.4)\times10^{-3}$~Mpc$^{-3}$ where we used Poisson statistics for the $1\sigma$ uncertainty. Thus the number density of LAEs around 6C0140 is comparable to that in the 1338 protocluster. 

The velocity distribution of the spectroscopically confirmed 1338 LAEs have a FWHM of $625\pm150$~km~s$^{-1}$. This is very narrow with respect to local galaxy clusters, but it is in agreement with the trend of decreasing velocity dispersion with increasing redshift \citep{venemans2007}. 

\begin{figure}
\resizebox{\hsize}{!}{\includegraphics{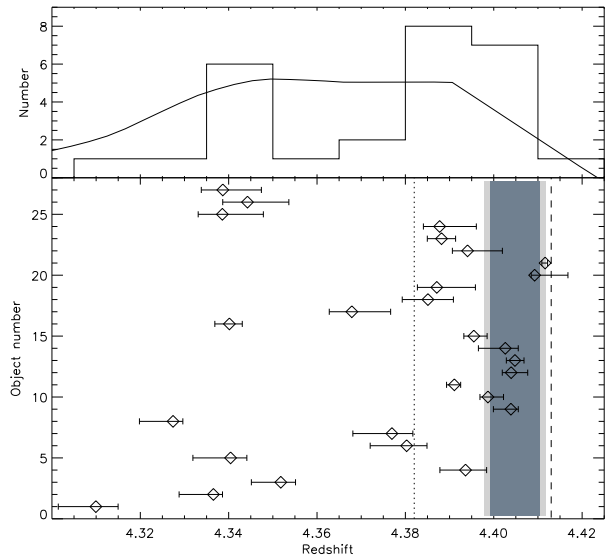}}
\caption{\label{fig:redzs} Lower panel: the estimated redshifts of the individual objects with their respective uncertainties. The dark shaded region represents the redshift interval the protocluster would have based on the 1338 velocity distribution, whereas the light shaded region takes into account the $1\sigma$ uncertainty. The protocluster region is chosen such as to maximise the number of protocluster candidates. The spectroscopic redshift of the radio galaxy ($z=4.413$) is marked by the vertical dashed line. The dotted line at $z\sim4.38$ indicates the approximate redshift limit set by the edge of the circular protocluster field as shown in Fig.~\ref{fig:spatial}. Upper panel: distribution of the estimated redshifts. Also shown is a curve that indicates how the effective selection area varies as function of redshift.}
\end{figure}

The estimated redshifts of the candidate LAEs are compared to the expected redshift range of the protocluster in the lower panel of Fig.~\ref{fig:redzs}. Note that the location of the protocluster region does not coincide with the redshift of the radio galaxy. Instead we have chosen the redshift of the protocluster such that the number of LAEs consistent with being in the protocluster is maximised. This results in an offset with respect to the radio galaxy of $\Delta z=0.0084$ or $\Delta v\sim465$~km~s$^{-1}$. This is consistent with what is observed in the 1338 protocluster where the 1338 radio galaxy is redshifted by 440~km~s$^{-1}$ ($\Delta z=0.0075$) with respect to the majority of the confirmed line emitters. Thus the radio galaxy does not have to be at the centre of the structure in redshift space. In the situation as shown in Fig.~\ref{fig:redzs} a total of 10 LAEs are consistent with being in the protocluster. This number decreases to 4 when we assume that the protocluster is centred on the radio galaxy. Note that a similar displacement of the radio galaxy with respect to the bulk of the galaxies is seen for the spatial distribution of both the 6C0140 and 1338 fields.  Both radio galaxies are located not at the centre of the spatial distribution of emitters, but more at the edge.  

Also shown in the upper panel of Fig.~\ref{fig:redzs} is the distribution of the estimated redshifts and a curve that indicates how the effective selection area varies as function of redshift. The area was estimated by determining the portion of the image for which $\lambda_{\rm low} < \lambda < \lambda_{\rm high}$ with $\lambda_{\rm low}$ and $\lambda_{\rm high}$ being respectively the lower edge of the TF1 tuning and the upper edge of the TF3 tuning. The curve indicates that, based on the effective selection area, we would expect the majority of the objects to have $4.35<z<4.39$. However, we find a disproportionately large number of objects at $z>4.38$ indicating that there is some concentration of LAEs close to the redshift of the radio galaxy. Applying a Kolmogorov-Smirnov test we determine that there is a probability of $4\times10^{-3}$ that the observed distribution is drawn from the expected distribution. The two distributions therefore differ at the $\sim3\sigma$ level.

The top panel of Fig.~\ref{fig:redzs} also shows that we are unable to observe objects that are located at $z>4.42$. This makes the reported overdensity of $8\pm5$ difficult to interpret. It may be that the protocluster structure extends beyond $z>4.42$. If this is the case, then the true overdensity may differ from the value presented here. Likewise, if the distribution of $\lambda_{\rm c}$ of the TFs had been chosen to probe larger redshift values, then such a blueshifted overdensity as found for 6C0140 may be underestimated or even missed altogether.

How do our results hold up when we account for the estimated contamination fraction discussed in Sect.~\ref{sec:interlopers}? In both the comparison with a blank field and the comparison with a $z\sim4.1$ protocluster we found 8 emitters in the 6C0140 field. Based on the minimum success rate of 75 per cent we therefore expect two interlopers in our `protocluster' sample. Redoing the comparison with a sample of six emitters the following results are obtained. In the comparison with the blank field of \citet{dawson2007} it is found that the 6C0140 field is denser by a factor of $7\pm4$. Thus the 6C0140 field harbours an overdensity of $\delta_{\rm g}=6\pm4$. The corrected number density in the 6C0140 field is $(2.5\pm1.3)\times10^{-3}$~Mpc$^{-3}$, which is also still in agreement with the 1338 field. The results presented here are therefore valid when contamination is taken into account and we conclude that the 6C0140 field is similarly overdense as the 1338 protocluster. This indicates that it may evolve into a massive galaxy cluster at $z=0$. Furthermore, this result supports the hypothesis that HzRGs are good tracers for galaxy overdensities in the early Universe.

\section{Conclusions \& outlook} \label{sec:conc}

We have presented the first search for high-$z$ protoclusters employing tunable narrowband filters. This pilot study focuses on the radio galaxy 6C0140+326 at $z\sim4.4$. Using a combination of three TF tunings we find a total of 27 unique LAEs in the field around 6C0140+326. Division of the field in a protocluster and a foreground field shows that the protocluster field contains significantly more objects than the foreground field. This indicates that there is a concentration of LAEs near the redshift of the radio galaxy.

A comparison to a blank field shows that the 6C0140 protocluster field contains an overdensity of $8\pm5$. The number density in the protocluster field is also comparable to that found in the 1338 protocluster at $z\sim4.1$. Both these results are robust when taking into account the possible presence of interlopers. 

With the availability of three separate TF tunings we also estimate the redshift distribution of the LAEs. Using results obtained for the 1338 protocluster we find that 4-10 of the LAEs have redshifts consistent with being in a redshift interval spanned by a typical $z\sim4$ protocluster. Also, the redshift distribution is different at the $3\sigma$ level from the expected distribution with a relatively large number of objects at $z>4.38$. This further strengthens the notion that there is a concentration of LAEs near the radio galaxy.

These results are further evidence that HzRGs pinpoint high density regions in the early Universe. The overdensity around 6C0140 may collapse at a later time to form a structure similar to a local galaxy group or cluster. Spectroscopic follow-up is needed to confirm this result.

We have shown that tunable filters are an excellent method of confirming the presence of protoclusters around HzRGs at any redshift. At the moment the wavelength range accessible to the red TF used in this study is limited to $\lambda>6500$~\AA~and therefore $z>4.3$. However, a blue TF covering the wavelength range $\lambda<6500$~\AA~will be commissioned in the near future. This will open up the redshift range $2< z <4$ which is where most of the known HzRGs are located. Our allocated GTC ESO large programme can then significantly expand the sample of protoclusters across cosmic time and this would allow an in-depth study of the evolution of these structures. 

\section*{Acknowledgements}
The authors would like to thank the anonymous referee for the useful comments that have helped in improving this paper. This work is based on observations made with the Gran Telescopio Canarias, installed in the Spanish Observatorio del Roque de los Muchachos of the Instituto de Astrof{\'i}sica de Canarias, in the island of La Palma. This work was partially supported by the Spanish Plan Nacional de
Astronom'a y Astrof'sica under grant AYA2008--06311--C02--01. EK acknowledges funding from Netherlands Organization for Scientific Research (NWO). NAH acknowledges support from STFC and the University of Nottingham Anne McLaren Fellowship. JK thanks the DFG for support via German-Israeli Project Cooperation grant STE1869/1-1.GE625/15-1.


\end{document}